\documentstyle[preprint,prl,aps]{revtex}
\newcommand{\degrees}{\mbox{$^{\circ}$}}
\newcommand{\piz}{\mbox{$\pi^{\circ}$}}
\newcommand{\pip}{\mbox{$\pi^{+}$}}
\newcommand{\rem}{\mbox{$R^{3/2}_{\sc EM}$}}

\begin{document}
\draft

\begin{flushleft}
{\bf Constraining the $(\gamma,\pi)$ amplitude for E2 ${\rm N}\rightarrow\Delta$ }
\end{flushleft}

The ratio of the E2/M1 ${\rm N}\rightarrow\Delta$ transitions provides an important discrimination
between nucleon structure models.  The quantity often compared to
model calculations is the ratio of the corresponding
photo-pion multipoles, $\rem=\Im m(E^{3/2}_{1+})/\Im m(M^{3/2}_{1+})$,
at the energy where the $\pi$N P$_{33}$ phase
goes through 90\degrees.  Recently, multipole analyses of two new data sets have
reported new values for this quantity, $-2.5 (\%) \pm 0.2(stat) \pm 0.2(sys)$
from Mainz \cite{beck}, and $-2.9 (\%) \pm 0.3(stat+sys) \pm 0.2(model)$
from LEGS \cite{legs}.  Here we investigate the puzzling fact that while these
multipole ratios appear to agree, the data from which they were extracted do
not.

At least 8
observables are necessary to specify the photo-pion amplitude
\cite{chiang}.  In the absence of such complete information, 
constraints from many observables are needed to avoid the
potential ambiguity discussed by Donnachie \cite{donna}
of higher partial wave strength appearing in lower partial
waves, and vice versa. The Mainz analysis of
\cite{beck} assumed complete dominance of S and P waves in a fit to only 2 observables. 
(In fact, the presence of an $s$-channel resonance guarantees non-Born $u$-channel contributions
to higher partial waves that cannot {\em a priori} be ignored.)
To examine the effect of these assumptions, we have repeated the analysis of the
Mainz data, successively adding constraints from other observables and
increasing the number of fitted partial waves. To avoid confusion, we have not included any LEGS
data.

In \cite{beck}, a fit of the Mainz 270-420 MeV $\sigma(\piz)$ cross section 
and $\Sigma(\piz)$ beam asymmetry 
data to powers of $\cos(\theta)$
was used to extract a quantity which, if only S and P waves contribute, should
be the ratio of {\piz} multipoles,
$R_{\piz} = \Re e[ E^{\piz}_{1+}\cdot(M^{\piz}_{1+}-M^{\piz}_{1-})^{*}]/
\{|E^{\piz}_{0+}|^{2} + |3E^{\piz}_{1+} - (M^{\piz}_{1+}
- M^{\piz}_{1-})|^{2}\}$.
The authors argued that $R_{\piz}(340)$ is
essentially \rem.  
We have performed an equivalent energy-independent analysis of
their data, fitting S and P-wave \piz multipoles, with phases fixed to the elastic
{\piz}p values of the VPI[SM95] $\pi$N scattering solution \cite{arndt}.
The result, $R_{\piz} = -2.31 (\%) \pm 0.54(stat+sys)$, 
is in agreement with the value of \cite{beck}.

We now add constraints from other data.
Following \cite{legs}, we use a K-matrix-like parameterization in an energy-dependent analysis of the
$(\gamma,\pi)$ multipoles with isospin $\tau$ and orbital angular momentum~$\ell$,
$A^{\tau}_{\ell\pm}(E_{\gamma}) = \{A^{\tau}_{B}(E_{\gamma}) + X(\varepsilon_{\pi})\}
(1 + iT^{\ell}_{\pi{\rm N}}) + \beta\cdot T^{\ell}_{\pi{\rm N}}$.
Here, $E_{\gamma}$ is the beam energy,
$A^{\tau}_{B}$ is the pseudo-vector Born multipole
(up to $\ell=19$), and $X$ is a polynomial in the $\pip$ kinetic
energy, $\varepsilon_{\pi}$.
VPI[SM95] values are used for the $\pi$N scattering
T-matrix, $T^{\ell}_{\pi{\rm N}}$ \cite{arndt}.

The isospin decomposition requires $\pip$ data. For this we use $\sigma(\pip)$ from 
\cite{fisher} and $\Sigma(\pip)$ from \cite{getman}. Varying S and P waves gives 
$-(3.27 \pm 0.35) \%$. In row 1 of the table we show the effect of extending the fit
to higher partial waves. The result, constrained by
only 2 observables, suffers from Donnachie's ambiguity \cite{donna} and is
not stable.

In row 2 we add constraints from other polarization ratios, the 
Bonn target (T) asymmetries \cite{dutz}, and
Khar'kov target and recoil (P) asymmetries \cite{getman,belyaev}.

The Mainz {\piz} cross sections of \cite{beck} are in very good agreement with the Bonn
results from \cite{genzel}.  However, the older Bonn data cover a larger angular range.
Including these in the fit gives the third row of the table.
The result is finally independent of the number of partial waves, but is significantly lower than
$R_{\piz}$ from \cite{beck}.

In conclusion, the procedure in \cite{beck} of fitting only S and P-wave 
multipoles to only 2 observables essentially corresponds to truncating an
expansion before it has converged.  
Other authors, refs. \cite{nimai,ron},
have disputed the accuracy of approximating $\rem$ with $R_{\piz}$. 
While we agree that this approximation is quite sensitive to the multipole solution, 
such inaccuracies are in fact
completely dwarfed by ambiguities in multipoles fitted with the Mainz procedure.
This work was supported by U.S. Dept. of Energy Contract No. AC02-76CH00016.

\begin{flushleft}
A.M. Sandorfi$^{1}$, J. Tonnison$^{2,1}$ and S. Hoblit$^{1}$\\
$^{1}$Physics Dept., Brookhaven National Lab, Upton, NY  11973\\
$^{2}$Physics Dept., Virginia Polytechnic Inst. and State Univ., Blacksburg, Va  24061
\end{flushleft}

\begin{flushleft}
Received  
\end{flushleft}

\begin{table}
\caption{Variation of {\rem} with increasing number of fitted partial waves, in successive
columns. Constraints from different observables between 270-420 MeV are increased in successive rows: 
row 1, refs. \protect\cite{beck,fisher,getman}; row 2, refs. \protect\cite{beck,fisher,getman,dutz,belyaev};
row 3: refs. \protect\cite{beck,fisher,getman,dutz,belyaev,genzel}. }
\label{tab1}
\begin{tabular}{cccc}
Data used in & \rem (\%) & \rem (\%) & \rem (\%) \\
analysis & $\ell_{\pi} = S - P$ & $\ell_{\pi} = S - D$ & $\ell_{\pi} = S - F$ \\
\tableline
\protect$\sigma, \Sigma$ &
$-(3.27 \pm 0.35)$ & $-(4.75 \pm 0.56)$ & $-(5.66 \pm 0.63)$ \\
\protect$\sigma, \Sigma, T, P$ &
$-(1.40 \pm 0.24)$ & $-(2.93 \pm 0.30)$ & $-(3.39 \pm 0.38)$ \\
\protect$\sigma, \Sigma, T, P, \sigma $ &
$-(0.58 \pm 0.08)$ & $-(0.71 \pm 0.10)$ & $-(0.58 \pm 0.12)$ \\
\end{tabular}
\end{table}

\end{document}